# A strategy for an accurate estimation of the basal permittivity in the Martian North Polar Layered Deposits

S. E. Lauro[1], G. Gennarelli[2], E. Pettinelli[1], F. Soldovieri[2], F. Cantini[3], A. P. Rossi[4], and R. Orosei[5]

[1] *Dipartimento di Matematica e Fisica, Università degli Studi Roma Tre, 00146 Rome, Italy*
[2] *Istituto per il Rilevamento Elettromagnetico dell'Ambiente, Consiglio Nazionale delle Ricerche, 80124, Naples, Italy*
[3] *École Polytechnique Fédérale de Lausanne (EPFL), Space Engineering Center (eSpace), CH - 1015 Lausanne*
[4] *Jacobs University Bremen, 28759 Bremen, Germany*
[5] *Istituto di Radioastronomia, Istituto Nazionale di Astrofisica, 40129 Bologna, Italy*

ABSTRACT

The paper deals with the investigation of the Mars subsurface by means of data collected by the Mars Advanced Radar for Subsurface and Ionosphere Sounding working at few MHz frequencies. A data processing strategy, which combines a simple inversion model and an accurate procedure for data selection is presented. This strategy permits to mitigate the theoretical and practical difficulties of the inverse problem arising because of the inaccurate knowledge of the parameters regarding both the scenario under investigation and the radiated electromagnetic field impinging on the Mars surface. The results presented in this paper show that, it is possible to reliably retrieve the electromagnetic properties of deeper structures, if such strategy is accurately applied. An example is given here, where the analysis of the data collected on Gemina Lingula, a region of the North Polar layer deposits, allowed us to retrieve permittivity values for the basal unit in agreement with those usually associated to the Earth basaltic rocks.

Key words—Mars exploration, ground penetrating radar, dielectric properties.

INTRODUCTION

Mars is today a cold, dry, and sterile world with a thin atmosphere made of $CO_2$. However, the geologic and compositional record of its surface reveals that in the past Mars had a thicker atmosphere and liquid water flowing on its surface (Jakosky and Phillips 2001). It has been postulated that life could have developed and that some primitive life forms may be existing even today (McKay and Stoker 1989). For this reason Mars Express, the first European Mars mission orbiting around the red planet since December 2003, has been designed to focus on two main issues: the inventory of ice or liquid water in the Martian crust and the search for possible traces of past or present biological activity on the planet (Chicarro 2004).

Mars Advanced Radar for Subsurface and Ionosphere Sounding (MARSIS) is the instrument on board Mars Express devoted to search for ice and water in the Martian subsurface. MARSIS antenna transmits radio pulses in Nadir direction, which are capable to penetrate the surface and be reflected by the dielectric discontinuities associated to structural or compositional changes (Picardi et al. 2004). In particular, MARSIS transmits 1 MHz bandwidth pulses centred at 1.8 MHz, 3 MHz, 4 MHz and 5 MHz, alternating the transmission at two different frequencies. The radar transmits a 10W, 250μs chirped (linear FM) pulse from a 40 m dipole antenna with a repetition frequency of 127 Hz. The choice of these relatively-low working frequencies makes MARSIS suitable for deep penetration in Mars subsurface, although the use of such frequencies entails a low range resolution, which is approximately 210 m in air after the application of a Hanning windowing to reduce the pulse side-lobes. MARSIS has been collecting data for over ten years acquiring thousands of radar images of the Martian subsurface (Orosei et al. 2014).

Despite the available huge amount of data, so far only a limited number of studies attempted to address the problem of supporting the identification of the compositional nature of the subsurface by quantitatively estimating the dielectric properties of the Martian subsoil. Indeed, the evaluation of such parameters from the backscattered signal requires to solve a non-linear inverse problem and different approaches have been proposed to retrieve such information from radar sounding data (e.g., Picardi et al. 2007; Campbell et al. 2008; Zhang et al. 2008; Grima et al. 2009; Carter et al. 2009; Lauro et al. 2010; Alberti et al. 2012; Lauro et al. 2012). In these works, the inversion algorithm is based on a simple model where a plane wave normally impinges on a layered medium with some additional assumptions (e.g., the complex permittivity of the first layer is known). The permittivity of the subsurface layer is retrieved from the ratio between the power of the electromagnetic field reflected at the Mars surface and the power of the electromagnetic field reflected by the unknown medium. Despite the simplicity of the inverse modeling, the estimation of the dielectric permittivity of the subsurface layers is a very challenging task, due to the uncertainties affecting the value of above mentioned power ratio and the inaccurate knowledge of the parameters involved in the inverse model.

In this paper, we develop a new strategy to address these problems by applying a procedure that ensures a stringent data selection with the aim to improve the reliability of the inversion scheme. As a test site, we chose Gemina Lingula, a region of Martian Northern polar cap characterized by a km-thick ice deposit consisting mostly of water ice with a small amount of dust and $CO_2$ (see e.g., Byrne 2009). We apply the inversion algorithm only to the data satisfying the imposed rigorous selection criteria in order to estimate the dielectric permittivity of the bedrock underlying the polar deposits. The values of the retrieved permittivity are found to be in good agreement with previous evaluations made in a different area of the Northern Polar cap of Mars (Lauro et al. 2010) and are also similar to those known for terrestrial basaltic rocks (Rust et al. 1999).

## INVERSION PROCEDURE AND DATA SELECTION

The overall data processing strategy consists of two steps: inversion scheme and data selection. For the clarity, these two logical steps are described in two separate subsections.

### Inversion Model

When MARSIS moves along an orbit, it collects a large number of observations: each observation consists of about 200 radar pulses (measurements) transmitted in a segment of orbit approximately 5-10 km long. Through unfocused SAR processing (azimuth and range compression), the measurements are reduced to a single radar trace (frame), which accounts for the power backscattered from the surface and subsurface discontinuities versus the two-way travel time. Thus, each frame probes an area on the Martian surface (footprint) that extends 5-10 km along track and 10-30 km across track, depending on the spacecraft altitude and surface roughness. As an example, Figure 1 shows the behaviour of the measured backscattered power versus two-way travel time for a single frame. The peak at time zero accounts for the reflection from the surface whereas the other main peak at about 18 μs is indicative of a reflection for a subsurface discontinuity.

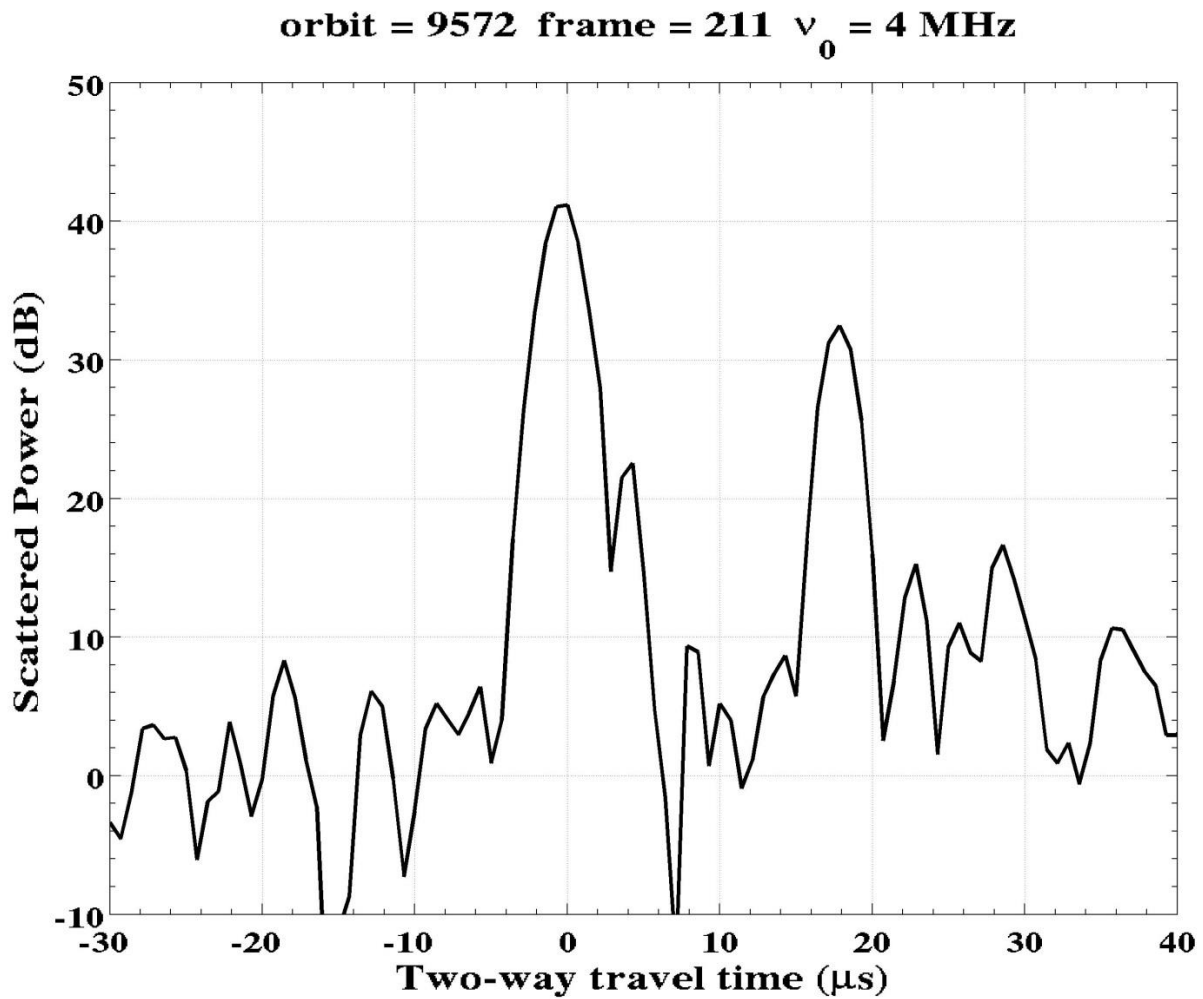

Figure 1 A typical MARSIS frame where the x-axis denotes the two-way travel and y-axis the scattered power in dB. The reference time is the surface echo arrival time and the reference power is the noise level.

The inversion model exploited for the quantitative estimation of the electromagnetic properties of the subsurface layers has been recently presented in (Lauro et al. 2010; Lauro et al. 2012) and is briefly described in the following. The input data for the inversion procedure is the ratio between the power $P_s$ reflected at surface by the air/soil interface and the power $P_{ss}$ reflected by the bedrock (subsurface discontinuity). This quantity is collected at each sounding point at the two operating frequencies of MARSIS. Note that here it is assumed that MARSIS only detects two sharp strong reflectors (surface and bedrock). The use of $P_s/P_{ss}$ as input parameter is dictated by the lack of accurate measurements of the field radiated by such a long dipole antenna (40 m). Indeed, the electromagnetic interaction of the antenna with the satellite platform would require the measurement of the field radiated by the antenna in operational conditions, which could not be performed due to the dimension and the characteristic of the antenna and the spacecraft.

A three-layered medium with flat interfaces is considered for the inversion model (see Figure 2); each layer is characterized by different electromagnetic parameters and the structure is illuminated by a plane wave with normal incidence (e.g., Lauro et al. 2010; Zhang et al., 2009).

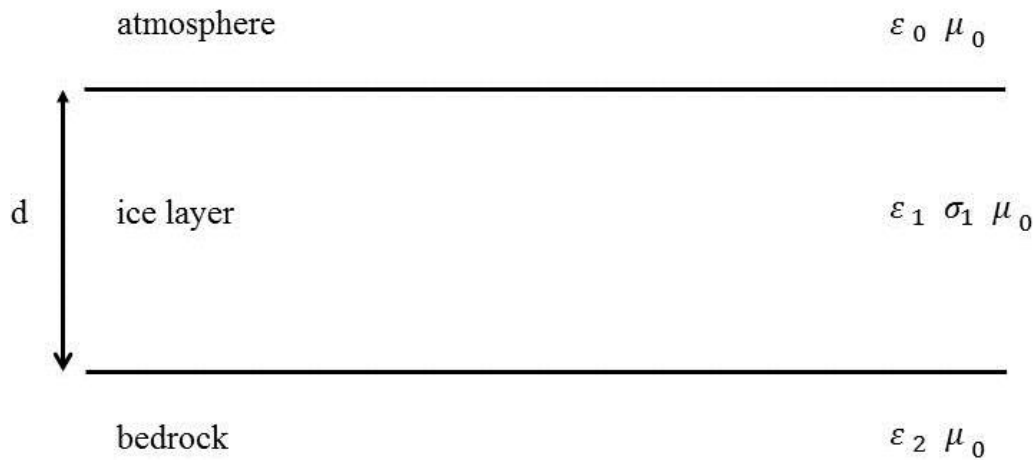

Figure 2 Geometry relevant to the inversion problem.

The upper layer (layer 0 above the Martian surface) is free-space with permittivity $\varepsilon_0$ and permeability $\mu_0$.

The second layer (layer 1 representative of the ice-layer) is assumed to be characterized by a known complex permittivity $\varepsilon_1 = \varepsilon_0\varepsilon_{1r} - j\sigma_1/(2\pi f)$ ($\epsilon_{1r}$ is the real part of the relative dielectric permittivity and $\sigma_1$ the electric conductivity). The third layer (layer 2 representative of the unknown subsurface layer, i.e., the bedrock) is assumed as a very low-loss non-magnetic medium with $\varepsilon_2 = \varepsilon_0\varepsilon_{2r}$ (Zhang et al., 2009). With regard to the electromagnetic features of layer 1, we make the assumption of a good dielectric (Balanis 1989), which holds when $\sigma_1/(2\pi f \varepsilon_0\varepsilon_{1r}) \ll 1$. Accordingly, the propagation constant in layer 1 can be approximated as $k_1 \approx 2\pi f \sqrt{\mu_0\varepsilon_0\varepsilon_{1r}} - j\sigma_1/2\sqrt{\mu_0/\varepsilon_0\varepsilon_{1r}}$.

In view of the above assumptions, the $P_s/P_{ss}$ ratio is given by

$$\frac{P_s}{P_{ss}} = \frac{|\Gamma_{10}|^2 \exp(4\alpha d)}{|\Gamma_{21}|^2 |1 - \Gamma_{10}^2|^2} \qquad (1)$$

where $\alpha = -\text{Im}[k_1]$ is the attenuation constant, $d$ is the thickness of layer 1, and

$$\Gamma_{ij} = \frac{\sqrt{\varepsilon_{jr}} - \sqrt{\varepsilon_{ir}}}{\sqrt{\varepsilon_{jr}} + \sqrt{\varepsilon_{ir}}} \qquad (2)$$

is the local reflection coefficient at the interface between the lower medium $i$ and upper medium $j$.

The thickness $d$ of the layer 1 (ice layer) is related to the two-way travel time $\Delta t$ between the first and second reflection of the electromagnetic signal according to the following equation:

$$d = \frac{\Delta t}{2\sqrt{\mu_0\varepsilon_0\varepsilon_{1r}}} \qquad (3)$$

From equation (1), we get a closed-form expression for the square amplitude of the reflection coefficient $\Gamma_{21}$, i.e.,

$$|\Gamma_{21}|^2 = \frac{P_{ss}}{P_s} \frac{|\Gamma_{10}|^2 \exp(4\alpha d)}{|1 - \Gamma_{10}^2|^2} \qquad (4)$$

Equation (4) is the basis for the inversion scheme and demands the knowledge of the $P_{ss}/P_s$ ratio as well as permittivity $\varepsilon_{1r}$ and conductivity $\sigma_1$ of layer 1, which are embedded in the local reflection coefficient $\Gamma_{10}$ and the attenuation term $\alpha$, respectively. The real part of the relative dielectric permittivity $\varepsilon_{1r}$ is assumed as a-priori information based on the knowledge of ice sheet properties (see next section); moreover, by taking into account the transparency of the ice in the North Pole (Grima et al. 2009) and considering a temperature of about 180 K for the ice at surface (Larsen and D. Dahl-Jensen 2000), we can assume in the $\sigma_1$ inversion algorithm which implies that $\alpha \cong 0$ (Pettinelli et al. 2015). The assumption of null attenuation (i.e., $\cong 0$) entails that the retrieved values of the dielectric permittivity of the basal units might be underestimated with respect to the true values (see equation (4)).

In light of the previous assumptions, the unknown quantity $\varepsilon_{2r}$ is estimated from the local reflection coefficient as

$$\Gamma_{21} = \frac{\sqrt{\varepsilon_{1r}} - \sqrt{\varepsilon_{2r}}}{\sqrt{\varepsilon_{1r}} + \sqrt{\varepsilon_{2r}}} \qquad (5)$$

Note that the layer 2 is expected to be a denser medium than layer 1 ($\varepsilon_{2r} > \varepsilon_{1r}$), as it should be composed by rock materials (Rust et al., 1999). Accordingly, the Fresnel reflection coefficient $\Gamma_{21}$ is a real and negative quantity. Finally, we can determine $\varepsilon_{2r}$ as

$$\varepsilon_{2r} = \varepsilon_{1r} \left(\frac{1 + |\Gamma_{21}|}{1 - |\Gamma_{21}|}\right)^2 \qquad (6)$$

Some consideration should be made about the robustness of the inversion scheme. As is known, an inversion problem is said to be "ill posed" when small uncertainties on the data produce large variations on the retrieved unknown (Bertero and Boccacci 1998), preventing the accurate estimation of the real value. Keeping in mind that the datum is the $P_{SS}/P_S$ ratio, which is linked to $\Gamma_{21}$ through equation (4), the degree of ill-posedness for the inverse problem at hand can be appreciated by evaluating the derivative of the reflection coefficient $\Gamma_{21}$ with respect to $\varepsilon_{2r}$, i.e.,

$$\frac{\partial \Gamma_{21}}{\partial \varepsilon_{2r}} = \frac{-\sqrt{\varepsilon_{1r}}}{\sqrt{\varepsilon_{2r}}(\sqrt{\varepsilon_{2r}} + \sqrt{\varepsilon_{1r}})^2} = -\frac{1}{\varepsilon_{1r} \, r_\varepsilon (r_\varepsilon + 1)^2} \qquad (7)$$

with $r_\varepsilon = \frac{\sqrt{\varepsilon_{2r}}}{\sqrt{\varepsilon_{1r}}}$.

Equation (7) clearly highlights that for high values of $\varepsilon_{2r}$, i.e., when $\varepsilon_{2r} \gg \varepsilon_{1r}$, i.e., $r_\varepsilon \gg 1$, $\partial \Gamma_{21}/\partial \varepsilon_{2r} \to$

0 thus there is poor "sensitivity" of the reflection coefficient $\Gamma_{21}$ with respect to $\varepsilon_{2r}$ variations. The plots reported in Figure 3 display the behavior of $\Gamma_{21}$ (left panel) and its derivative (right panel) versus relative permittivity $\varepsilon_{2r}$, which is supposed to vary in the range of 3-80. Here, the real part of the dielectric permittivity of the ice-layer $\varepsilon_{1r}$ is assumed equal to 3.1, according to previously published data (Grima et al. 2009). As expected, $\Gamma_{21}$ assumes negative values with its amplitude growing monotonically with respect to $\varepsilon_{2r}$. Most notably, the derivative of the local reflection coefficient $\Gamma_{21}$ is different from zero for low values of $\varepsilon_{2r}$ (<15) whereas it becomes almost equal to zero for higher values (see right panel of figure 3), thus making the inversion less robust with respect to measurement fluctuations and uncertainties.

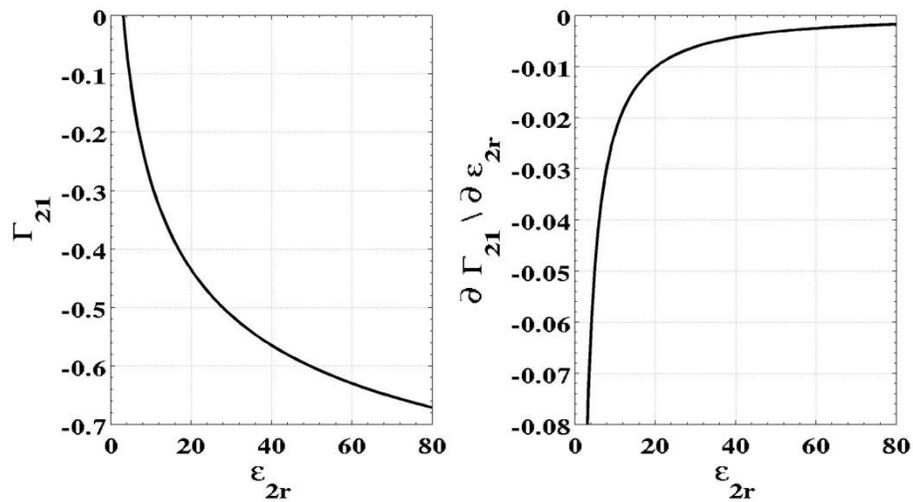

Figure 3 Local reflection coefficient $\Gamma_{21}$ (left panel) and its derivative (right panel) versus relative dielectric permittivity $\varepsilon_{2r}$.

**Data selection**

The reliability of the inversion scheme requires an accurate selection of the radar observations to fulfil the assumptions introduced by the model described in Inversion Model Subsection. The criteria applied for the data selection are presented below.

*The first criterion* deals with the smoothness of Martian surface in the investigated area, as the inverse model requires that the surface is flat with respect to the wavelength of the radar signal. To satisfy such criterion, we have chosen a roughness threshold of $\lambda/16$ (with $\lambda$ being the free-space wavelength) and we have selected only the frames acquired in areas where the roughness was smaller than the threshold. For the frequencies 1.8, 3, 4, 5 MHz, the maximum admissible surface roughness (root mean square height) are 10.4, 6.25, 4.7, 3.7 m, respectively. Note that the surface roughness was independently estimated by resorting to the data acquired by Mars Orbiter Laser Altimeter (MOLA), whose primary objective is to map globally the topography of Mars at a level suitable for geophysical, geological, and atmospheric circulation studies of Mars. This altimeter operates at a wavelength of 1.064 µm and with a range resolution of 37.5 cm and provides information about the topography with a vertical accuracy of 1m (Smith et al. 2001).

*The second criterion* refers to the assumption of normal incidence for the plane wave impinging the surface. This assumption is consistent with the observation modality of MARSIS that looks in Nadir direction. Since, as expected, the Martian surface is subjected on slope variations, we have selected only the data collected where the plane wave field was impinging with an angle smaller than 0.5 degrees with respect to the normal to the surface.

*The third criterion* deals with the actual signal to noise ratio (SNR) on the data. The SNR is computed as the ratio between the peak of the reflected power at the air/soil interface and the noise power estimates as the average of the first portion of the signal before the first peak, i.e., the arrive of the surface echo. We have selected only data frames characterized by SNR ≥35 dB.

*The fourth criterion* regards the rejection of those frames affected by ionospheric distortions. Indeed, MARSIS data are corrected for such distortion through the application of a proper signal processing algorithm (Safaeinili et al. 2003). To select only the frames that do not show any ionospheric effect, we adopted a criterion based on the similarity between the original data frames (i.e., before the ionospheric correction) and the processed ones. The similarity was quantified in terms of the maximum of the cross-correlation function

$$C = max_\tau \int r_c(t) r_{nc}(t-\tau) dt \qquad (8)$$

where $r_c(t)$ and $r_{nc}(t)$ are the corrected and non-corrected radar echoes, respectively. Note that $C$ has maximum value equal to 1 when $r_c(t)$ and $r_{nc}(t)$ have the same shape and only differ for a temporal shift.

We selected the frames having $C \geq 0.95$.

## DATA ANALYSIS AND RESULTS

The strategy presented in the previous section (data selection and inversion algorithm) has been applied to the radar data collected in the Gemina Lingula region (the area delimited by the white line in Figure 4), which is located within the Martian Northern polar cap. As mentioned in the Introduction, this region is characterized by a km-thick ice deposit about 250000 km$^2$ in extension, consisting mostly of water ice with a small amount of dust and $CO_2$ (Byrne 2009). Figure 4 also illustrates Gemina Lingula topography measured by MOLA.

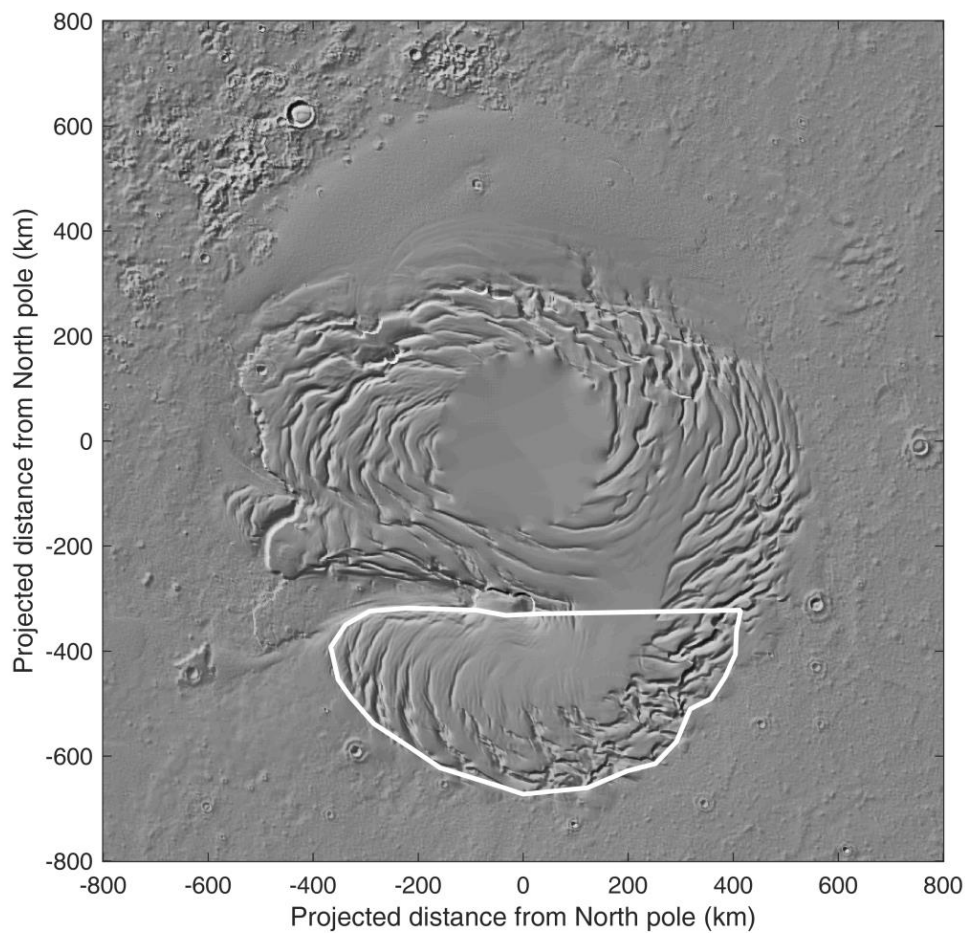

Figure 4 Shaded relief map of the Northern polar cap of Mars based on the MOLA topographic dataset (Smith et al. 2001). The white contour outlines Gemina Lingula.

The data analysis has been applied to 705 orbits passing over Gemina Lingula that corresponded to a total number of 73770 frames. Note that large part of the frames has been acquired at 3 and 4 MHz and only few data were collected at 1.8 and 5 MHz. The selection of the radar traces suitable for the inversion has been carried out in agreement with the four criteria previously described regarding surface roughness, incidence angle, SNR, and ionospheric effects.

Furthermore, two additional criteria have been introduced on the basis of the peculiar topographic features of the Gemina Lingula region. As clearly visible in Figure 4, the area under investigation is characterized by the presence of several superficial channels, which can generate strong radar echoes arriving after the surface echo. As such echoes can mask, or be mistaken for, subsurface echoes, numerical electromagnetic models of surface scattering have been developed (see e.g., Nouvel et al. 2004; Spagnuolo et al. 2011) to validate the detection of subsurface interfaces in MARSIS data. They are used to produce simulations of surface echoes, which are then compared to the ones detected by the radar: any secondary echo visible in radargrams but not in simulations is interpreted as caused by subsurface reflectors.

We developed a code for the simulation of radar wave surface scattering (Cantini et al. 2014) based on the algorithm of Nouvel et al. (2004). The MOLA topographic dataset (Smith et al. 2001) was used to represent the Martian surface as a collection of flat plates called facets. Radar echoes were computed as the coherent sum of reflections from all facets illuminated by the radar. The computational burden of simulations was well above the capabilities of PC's, and required the use of a supercomputer. The selection has thus been carried out by comparing real and simulated frames and discarding all those frames in which simulations showed secondary surface echoes less than 10 dB weaker than nadir echoes.

Moreover, on the basis of the composition and thickness of the polar deposits (Grima et al. 2009), we have only selected frames for which the two-way travel time $\Delta t$ was larger than 12 µs. The choice of this travel time was made in order to consider only the subsurface discontinuities at a depth larger than 1 km (see equation (3), where the relative dielectric permittivity of the ice-layer is assumed equal to 3.1); this assumption is consistent with the a-priori information about the thickness of the ice-layer in Gemina

Lingula area (Grima et al. 2009).

Table I summarizes the values of the parameters adopted for the selection of the frames. Such severe selection drastically reduced the number of useful data: only 124 out of the original 73770 frames can be considered suitable for the inversion. Most of these data refer to the acquisitions at 3 and 4 MHz, since the 1.8 MHz frames have been completely removed and the number of the 5MHz frames significantly decreased.

TABLE I

PARAMETERS ADOPTED FOR FRAME SELECTION

| SELECTION CRITERION | VALUE |
|---|---|
| Max. incidence angle [°] | 0.5 |
| Max. surface roughness | $\lambda/16$ |
| Min. SNR [dB] | 35 |
| Max. cross-correlation | 0.95 |
| Min. two-way travel time [µs] | 12 |
| Min. simulated $P_s/P_{ss}$ [dB] | 10 |

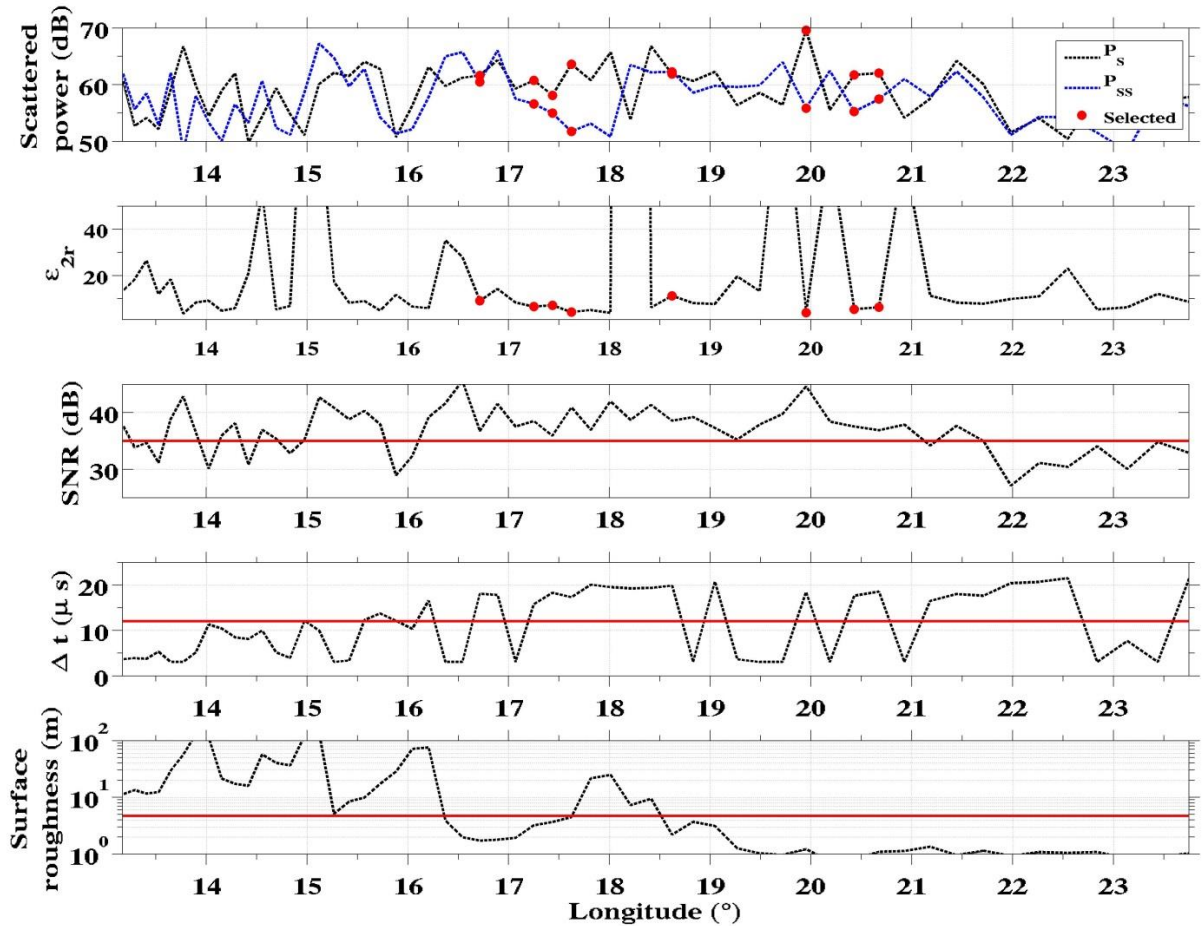

Figure 5 An example of data selection for the orbit 11905 at 4 MHz as a function of longitude. The surface and subsurface powers are plotted (dashed lines) in the first panel together with the selected values for the inversion procedure (solid circles). The estimated subsurface permittivity (dashed lines) and the selected results (solid circles) are plotted in the second panel. In the lower panels are plotted the SNR, two-way travel time and the surface roughness; the red lines represent the values of the parameters adopted in the data selection (see table 1).

An example of the effect of the selection criteria is shown in Figure 5 for the orbit 11905 at 4 MHz. The power $P_s$ reflected at surface and the power $P_{ss}$ reflected at subsurface are plotted in the first panel as dashed lines; the solid circles indicate the selected values after the adoption of the criteria in Table 1. The second panel reports the values of subsurface permittivity estimated applying equations (4-6) using the whole data set (dashed line) and the selected frames (solid circles). In the lower panels are plotted the SNR and the two-way travel time values extracted from the radar data and the surface roughness values from MOLA data; the red lines represent the adopted values of parameters applied in the data selection. For brevity, the other three criteria are not shown in Figure 5 as, in this specific case, they do not further filter the data. Figure 5 shows how the severe data selection is able to filter out very high and unrealistic values of the dielectric permittivity (in this specific case it results 12 permittivity values greater than 20).

The first main result obtained after applying the selection strategy proposed above, regards the surface radar response in the investigated area. In fact, thanks to the high resolution of MOLA data, the simulator has also been able to estimate the effect of the roughness on the power $P_S^{Sim}$ reflected by the surface. It should be highlighted however that the power of the simulated signals is only affected by the surface roughness, whereas the power of MARSIS data can also be influenced by local variations of the surface composition (i.e., dielectric properties). Thus, in principle, comparing the statistical distribution of the real $P_S$ and simulated $P_S^{Sim}$ power for the subset of data satisfying the first four criteria (as the last two specifically refer to subsurface features), we should be able to infer some information on the compositional homogeneity of the investigated area. Figure 6 illustrates the results of this analysis in terms of power distribution normalized to the average power value proper of each frequency. In the figure, left panels refer to measured data and right panels to simulated ones. Note that due to the strong reduction of data set after selection, only channels 3 and 4 MHz are shown as they preserve a meaningful number of data frames. Figure 6 shows that the four distributions exhibit similar width (about 10 dB) and do not significantly differ from each other; this result suggests that the strong fluctuation in the surface power can mainly be ascribed to the roughness, since every local variation in dielectric property is masked by such strong first order effect.

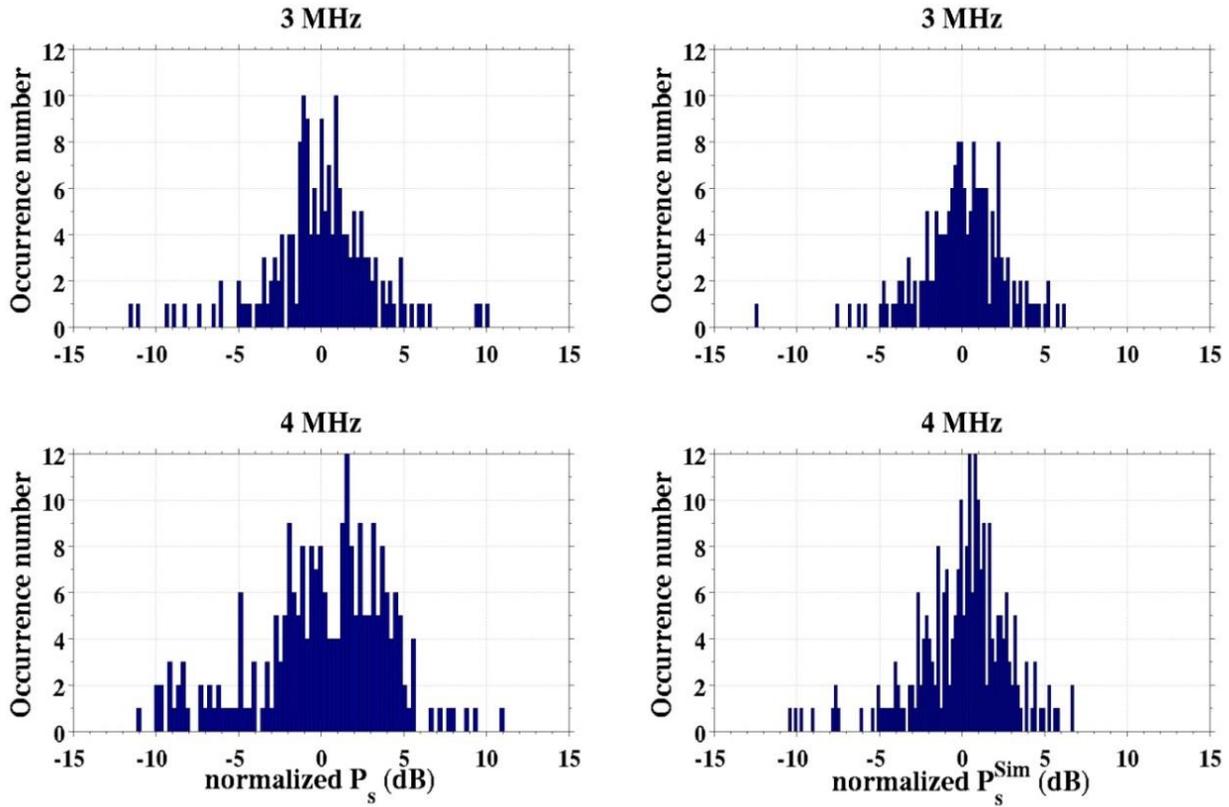

Figure 6 Histograms of measured (left panels) and simulated (right panels) surface reflected power (expressed in dB) normalized to their average value for 3 and 4 MHz frequency.

The latter consideration allows us to assume a spatial homogeneity for the material composing Gemina Lingula area. Thus, we have imposed a relative dielectric permittivity value of 3.1 for $\varepsilon_{1r}$ (i.e. the permittivity value of pure water ice) in the inversion algorithm (equations 4 and 6); this choice is based on the results reported in (Grima et al. 2009).

Figure 7 depicts the spatial map of the estimated subsurface permittivity $\varepsilon_{2r}$, where the small circles represent the real footprint of MARSIS antenna over the surveyed region and the color code (ranging between 3 and 10) the estimated value of the relative dielectric permittivity of the bedrock. The 124 dots are randomly distributed and no specific clustered zone is recognizable in the investigated area. Moreover, large part of the dots is blue-green indicating that the bedrock retrieved permittivity values are predominantly lower than 6. The average value of the bedrock permittivity can be extracted from the histogram shown in Figure 8, which confirms that the distribution is peaked around 5 and most of the estimated values range between 4 and 8. These values are in very good agreement with previous findings (Lauro et al. 2010) and are compatible with the permittivity of terrestrial basaltic rocks (Rust et al. 1999).

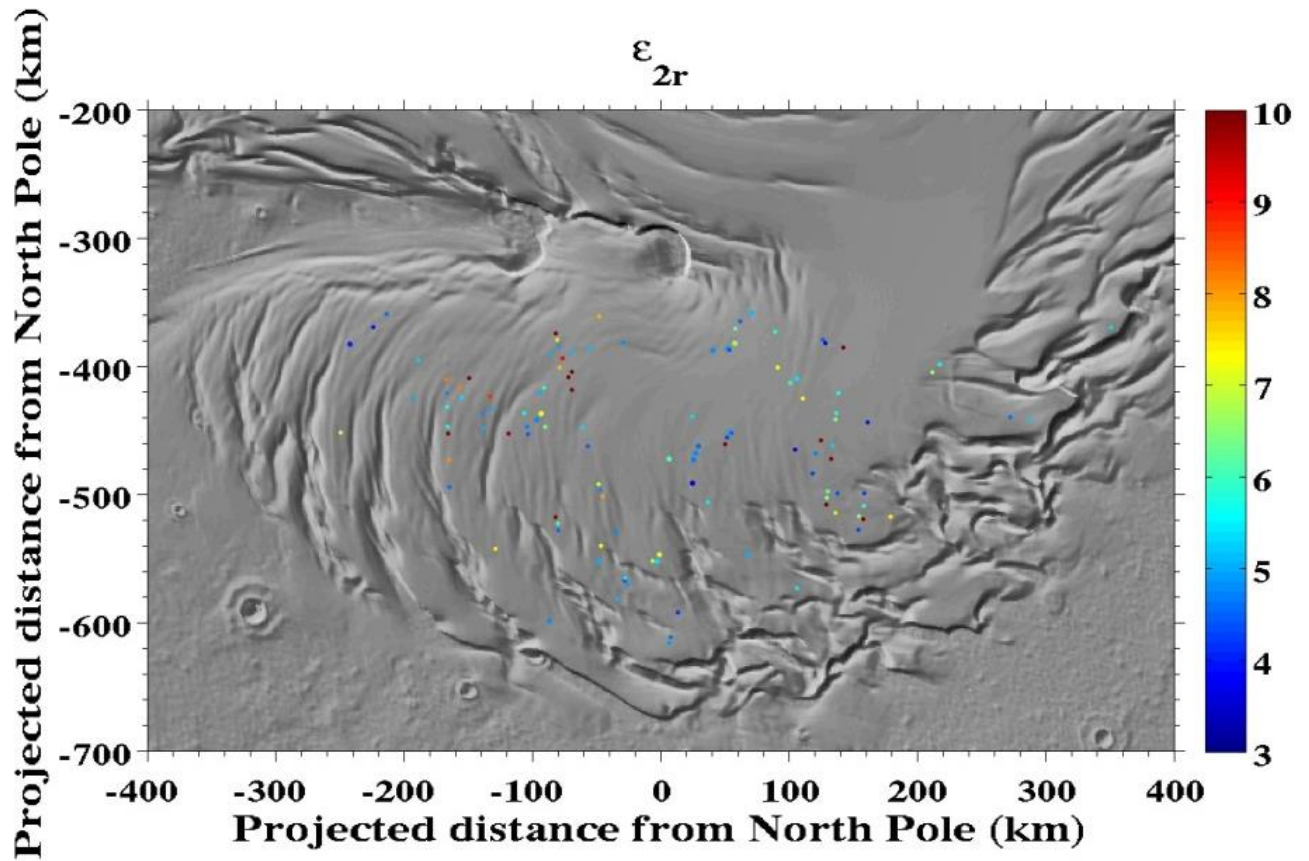

Figure 7 Reconstruction of the relative dielectric permittivity of the subsurface in the range 3-10 based on a relative dielectric permittivity of the first layer equal to 3.1. The small circles denote the real footprint of MARSIS.

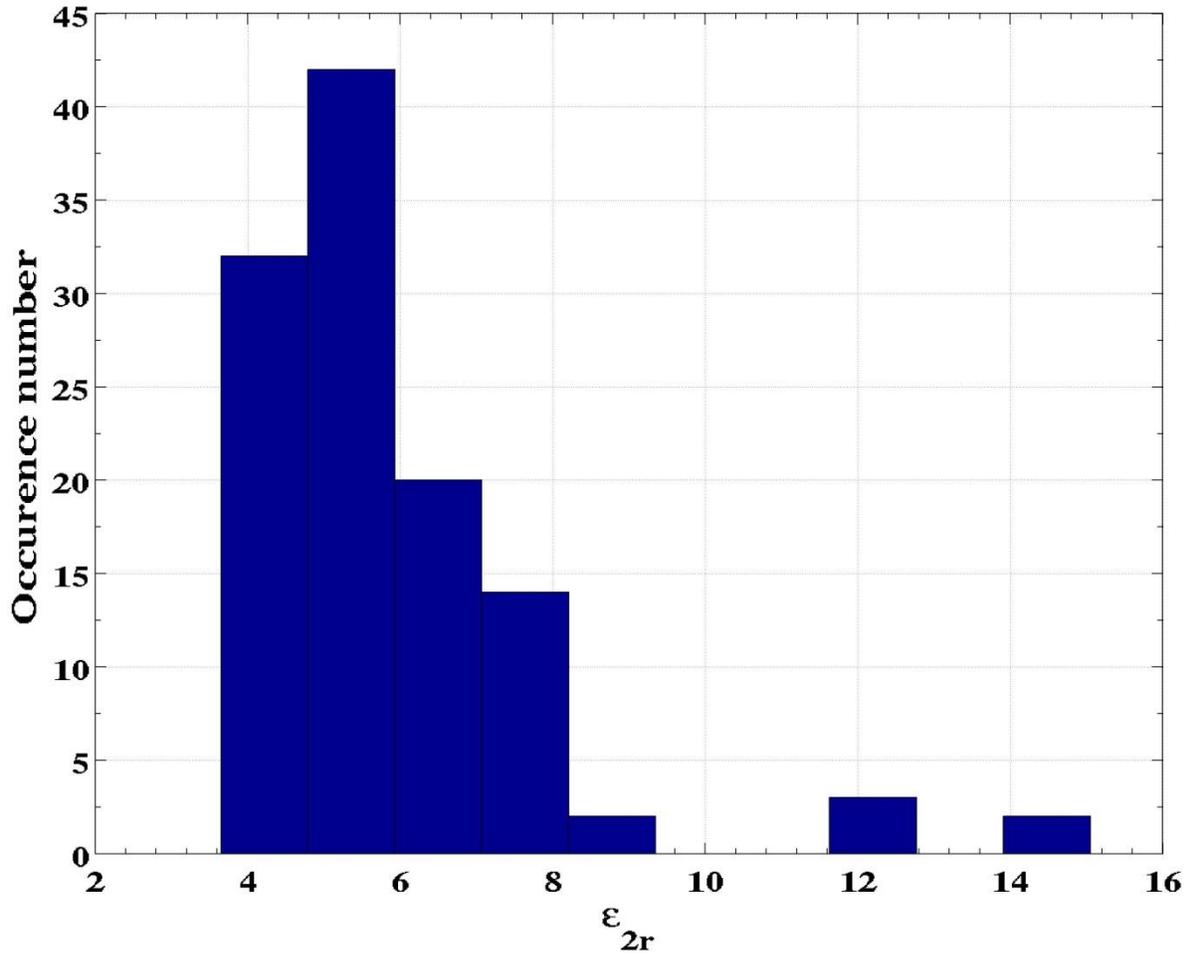

Figure 8 Histogram of the estimated relative subsurface permittivity based on a constant relative permittivity of the first layer equal to 3.1.

**CONCLUSIONS**

In this work we presented a novel strategy to apply a simple inversion algorithm for Martian subsurface dielectric properties estimation using MARSIS observations. We have shown that, since the data cannot be calibrated and are affected by several factors peculiar to Mars environment (ionosphere, surface topography and roughness, noise, etc.), it is mandatory to operate a stringent data selection before using any inversion procedure. Our procedure has been tested on a "well known" area, as Gemina Lingula, which has been studied for a long time and exhibits several favorable conditions in terms of composition, topography and roughness. For this large area, we considered a starting dataset of more than seventy thousand observations and, after the rigorous selection, we drastically reduced the number of reliable radar frames useful for the inversion to about a hundred. Nevertheless the remaining data, even if scattered in the investigated area, are quantitatively consistent and have been used to estimate the average permittivity of the bedrock in the overall region.

The approach proposed here requires a very large dataset to ensure a sufficient number of outcomes; as a consequence it is mainly suitable for the investigation of fairly large areas. In smaller regions, although the possibility to obtain a significant statistics is low, such strategy could still help to define zones where the outcomes are spatially clustered and quantitatively consistent.

**ACKNOWLEDGMENTS**

This work was supported by the Italian Space Agency (ASI) through contract no. I/032/12/1. The numerical code for the simulation of surface scattering was developed at the Consorzio Interuniversitario per il Calcolo Automatico dell'Italia Nord-Orientale (CINECA) in Bologna, Italy. Simulations were produced thanks to the Partnership for Advanced Computing in Europe (PRACE), awarding us access to the SuperMUC computer at the Leibniz-Rechenzentrum, Garching, Germany. This research has made use of NASA's Astrophysics Data System.